\documentclass[aps,twocolumn,preprintnumbers,nofootinbib,superscriptaddress]{revtex4}
\usepackage{eurosym}
\usepackage{amsmath}
\usepackage{graphicx}
\usepackage{amsfonts}
\usepackage{array}
\usepackage{amsthm}
\usepackage{bm}
\usepackage{palatino}
\usepackage{mathpazo}
\usepackage{supertabular}
\usepackage{subfig}

\usepackage[colorlinks]{hyperref}
\usepackage{subfig}
\usepackage[font=small,labelfont=bf,justification=justified,format=plain]{caption}
\usepackage[font=small,labelfont=bf,justification=justified]{caption}
\usepackage{color}
\usepackage{booktabs}
\usepackage{multirow}
\usepackage[labelfont=bf,labelsep=quad,justification=justified]{caption}
\usepackage{xcolor}
\newcommand{\be}{\begin{equation}}
\newcommand{\ee}{\end{equation}}
\newcommand{\ba}{\begin{eqnarray}}
\newcommand{\ea}{\end{eqnarray}}
\newcommand{\bal}{\begin{align}}
\newcommand{\eal}{\end{align}}
\newcommand{\lb}{\label}

\newcommand{\bw}{\begin{widetext}}
\newcommand{\ew}{\end{widetext}}

\begin{document}

\title{Quasinormal Modes of Black Holes Surrounded by Dark Matter and Their Connection with the Shadow Radius}
\author{Kimet Jusufi}
\email{kimet.jusufi@unite.edu.mk}
\affiliation{Physics Department, State University of Tetovo, Ilinden Street nn, 1200,
Tetovo, North Macedonia}
\affiliation{Institute of Physics, Faculty of Natural Sciences and Mathematics, Ss. Cyril
and Methodius University, Arhimedova 3, 1000 Skopje, North Macedonia}

\begin{abstract}
The purpose of this article is twofold. First, we highlight a simple connection between the real part of quasinormal modes (QNMs) in the eikonal limit and shadow radius of BHs and then explore the effect of dark matter on the QNMs of massless scalar field and electromagnetic field perturbations in a black hole (BH) spacetime surrounded by perfect fluid dark matter (BHPFDM). Using the WKB approximation we show that the quasinormal mode spectra of BHPFDM deviate from those of Schwarzschild black hole due to the presence of the PFDM encoded by the parameter $k$.  Moreover it is shown that for any $k>0$, the real part and the absolute value of the imaginary part of QNM frequencies increases and this means that the field perturbations in the presence of PFDM decays more rapidly compared to Schwarzschild vacuum BH. We point out that there exists a reflecting point $k_0$ corresponding to maximal values for the real part of QNM frequencies.  Namely, as the PFDM parameter $k$ increases in the interval $k<k_0$, the QNM frequencies increases and reach their maximum values at $k=k_0$. Finally we show that $k_0$ is also a reflecting point for the shadow radius while this conclusion can be deduced directly from the inverse relation between the real part of QNMs and the shadow radius. 
\end{abstract}
\maketitle

\section{Introduction}

One of the most exciting predictions of Einstein's general theory of relativity are black holes.  Physicists have been studying black holes for decades but in the same time black holes have been a subject of debate for a long time. However, nowadays, the situation has changed drastically due to the recent announcements  of the detection of
gravitational waves (GWs) of black hole binary mergers \cite{AbbottBH} by
the LIGO and VIRGO observatories and the captured image of the black hole shadow of a
supermassive M87 black hole by the Event Horizon Telescope collaboration \cite%
{Akiyama1,Akiyama4}. Alternatively,  there are other indirect methods to infer their existence, for example by observing high-energy phenomena such as X-ray emission and jets, and the motions of nearby objects in orbit around the hidden mass.

On the other hand, the discovery of gravity waves has opened a new window in our understanding of the Universe. Future observations of gravity waves can be used to test many alternative theories of general relativity.  The evolution of binary black holes is conventionally split into three stages: inspiral, merger and the ring down. During the inspiral phase the signal provides characterizes of the masses and the spins of compact objects and can be described by the post-Newtonian
approximation \cite{Blanchet}. The merger phase occurs after the
inspiral phase with a rapid collapse of the two objects to form a black
hole and can ne studied using numerical relativity  \cite{Pretorius,Campanelli,Baker}. The
ringdown phase is the final stage and describes a perturbed black hole that emits GWs in the form of
quasinormal radiation \cite{BertiCardosoWill}. In this way, the total evolution is described by a combining the numerical and the analytical methods. The perturbation theory of Schwarzschild black hole and its stability under small perturbations was studied in Refs. \cite{Regge,Zerilli}. In a short period of time the frequencies of perturbations or also known as
the QN frequencies (QNFs)  have been investigated by using other analytic and numerical methods
\cite{Mashhoon}-\cite{Turimov:2019afv}. On the other hand, the shadow of a Schwarzschild black hole was first studied by Synge \cite{Synge66} and Luminet \cite{Luminet79} and the same for Kerr black hole was studied by Bardeen \cite{DeWitt73}. Since then various authors have studied shadows in modified theories of gravity and wormholes \cite{Hou:2018avu}-\cite{Dokuchaev:2019jqq}. 

Our aim in this paper is to explore the effect of perfect fluid dark matter on two types of field perturbation: scalar and electromagnetic  perturbations. Toward this goal we are going to use the black hole solution surrounded by PFDM recently proposed in Refs. \cite{Li,K1} along with the sixth order WKB approximation developed by Konoplya \cite{KonoplyaWKB}. It was shown by Cardoso et al. \cite{cardoso} that the real part of the QNMs is related to the angular velocity of the last circular null geodesic while Stefanov et al. \cite{Stefanov:2010xz}  found a connection between black-hole quasinormal modes in the eikonal limit and lensing in the strong deflection limit. With these results in mind we shall explore the connection between the QNMs and the shadow radius.

This paper is organized as follows. In Section II, we study the connection between QNMs in the eikonal regime and the BH shadow. In Section III we use the WKB approximation to study the QNMs of scalar fields in spacetime of a BHPFDM. In Section III, we investigate the QNMs of the electromagnetic field.  In Section IV we study the greybody factors. In Section V, the connection between QNMs and the PFDBH shadow radius. In Section VII, we comment on our results.

\section{Connection between shadow radius and QNMs}
Let us start by writing a static and spherically symmetric black hole solution
\begin{equation}
ds^2=-f(r)dt^2+\frac{dr^2}{f(r)}+r^2\left(d\theta^2+\sin^2\theta \,d\phi^2\right),
\end{equation}
to analyze the evolution of the photon in the above spacetime metric. In order to study the the null geodesics in the above black hole spacetime  one has to use the Hamilton-Jacobi equation given by
\begin{equation}
\frac{\partial S}{\partial \lambda} = - \frac{1}{2}g^{\mu\nu} \frac{\partial S}{\partial x^\mu} \frac{\partial S}{\partial x^\nu},
\end{equation}
with $\lambda$ being the affine parameter of the null geodesic and $S$ is the Jacobi action. It is well known that the Jacobi action $S$ can be separated as follows
\begin{equation}
S = \frac{1}{2} m^2 \lambda - E t + L \phi + S_{r}(r) + S_\theta (\theta),
\end{equation}
with $m$ being the mass of the particle moving and for the photon one has to set $m=0$. Furthermore $E$ and $L$ are the energy and angular momentum of the photon, respectively. In addition, the functions $S_r(r)$ and $S_\theta(\theta)$ depend only on $r$ and $\theta$, respectively. It is straightforward to show the following four equations of motions describing photons
\begin{eqnarray}
\frac{dt}{d\lambda} &=& \frac{E}{f(r)}, \\
\frac{dr}{d\lambda} &=& \frac{\sqrt{R(r)}}{r^2}, \\
\frac{d\theta}{d\lambda} &=& \frac{\sqrt{\Theta(\theta)}}{r^2}, \\
\frac{d \phi}{d \lambda} &=& \frac{L \csc^2\theta}{r^2}.
\end{eqnarray}
with
\begin{eqnarray}
R(r) &=&  E^2r^4 - (\mathcal{K}+L^2) r^2 f(r), \\
\Theta(\theta) &=& \mathcal{K} - L^2 \csc^2 \theta \cos^2\theta.
\end{eqnarray}
t this point it convenient to express the radial geodesics in terms of the effective potential $V_{\rm eff}(r)$ as follows
\begin{equation}
 \left(\frac{dr}{d\lambda}\right)^2 + V_{\rm eff} (r)= 0,
\end{equation}
with
\begin{equation}
V_{\rm eff}(r) = - r^{-4}R(r)/E^2 = -1 + \frac{f(r)}{r^2} (\xi^2 +\eta),
\end{equation}
where we have introduced
\begin{equation}
\xi = \frac{L}{E},\;\;\eta = \frac{\mathcal{K}}{E^2}.
\end{equation}
The motion of the photon can be determined by these two impact parameters. To determine the geometric sharp of the shadow of the black hole, we need to find the critical circular orbit for the photon, which can be derived from the unstable condition
\begin{equation}\lb{condition}
R(r)=0,\;\; \frac{dR(r)}{dr} =0 ,\;\;\; \frac{d^2 R(r)}{dr^2} >0.
\end{equation}
Using the above conditions (\ref{condition}) for our spherical symmetric black hole one can show that
\begin{equation}\lb{PSradius}
2 - \frac{rf'(r)}{f(r)}=0.
\end{equation}
By solving this equation one can determine the radius of the photon sphere $r_{\rm ps}$. In particular can write $\xi^2+\eta$ in the following form
\begin{equation}
\xi^2 +\eta = \frac{r^2_{\rm ps}}{ f(r_{\rm ps})}.
\end{equation}

On the other hand, the observable $R_{\rm s}$, represents the size of the shadow of the black holes which can be expressed via celestial coordinates $(\alpha,\beta)$ by the simple relation
\begin{equation}\lb{RS}
R_{\rm s} = \sqrt{\alpha^2+\beta^2} =\frac{r_{\rm ps}}{ \sqrt{f(r_{\rm ps})}}.
\end{equation}

Another observable quantity is the angular radius of the shadow which can be defined in terms of the shadow radius and the distance from the black hole as seen by an observer located far away
\begin{equation}\lb{thetas}
\theta=R_s/D.
\end{equation}
It was shown by Cardoso et al. \cite{cardoso} that the real part of the QNMs in the eikonal regime is related to the angular velocity of the last circular, null geodesic, and the imaginary part was related to the Lyapunov exponent that determines the instability time scale of the orbit \cite{cardoso}
\begin{equation}
\omega_{QNM}=\Omega_c l -i \left(n+\frac{1}{2}\right)|\lambda|.
\end{equation}
This result was proved to be valid for not only the static, spherical spacetime, but also the equatorial orbits in the geometry of rotating black holes solution. On the other hand, Stefanov et al. \cite{Stefanov:2010xz}  found a connection between black-hole eikonal quasinormal modes, i.e. it states the relation between the
geodesics and quasinormal spectrum  which is valid at high multipole numbers only and lensing in the strong deflection limit [here we shall introduce temporary the speed of light]
\begin{equation}
\Omega_c=\frac{c}{\theta D_{OL}},\,\,\,\lambda=\frac{c\,\ln \tilde{r}}{2 \pi \theta D_{OL}},
\end{equation}
where $D_{OL}$ is the distance between the observer and the lens and $\theta$ is the angular position of the image that is closest to the black hole, $\lambda$ is the Lyapunov exponent and determines the instability time scale. Actually, even for high multipoles $l$, the correspondence is not guaranteed 
for every type of fields, but only for the test fields, as was shown by  
Konoplya and Stuchlik \cite{Konoplya:2017wot}. Now by making use of the relations (17),  (18) and (19) we find that in eikonal regime the real part of QNMs is inversely proportional to the shadow radius
\begin{equation}
\omega_{\Re } \propto \frac{1}{R_S},
\end{equation}
where we have identified $D_{OL}$ with $D$. Hence the real part of the QNMs and shadow radius are related by the simple equation
\begin{equation}
\omega_{\Re} = \lim_{l \gg 1} \frac{l}{R_S},
\end{equation}
which is accurate only in the eikonal limit having large values of $l$. Again, this correspondence is not guaranteed for gravitational fields, as the link between the null geodesics and
quasinormal modes is violated in the Einstein-Lovelock theory even in the eikonal limit \cite{Konoplya:2017wot}. In  Ref. \cite{Wei:2019jve} authors pointed out an equivalent expression to (21) relating the angular velocity with the shadow radius for Schwarzschild BH.  In the present paper, we explicitly relate the real part of QNMs in the eikonal limit with the shadow radius. This connection between QNMs and shadow radius is a reflection of the fact that the gravitational waves are treated as massless particles propagating along the last null unstable and slowly leaking out to infinity. 
In addition, we can express also the Lyapunov exponent in terms of shadow radius $R_S$ and flux ratio $\tilde{r}$ reads 
\begin{equation}
\lambda=\frac{\ln \tilde{r}}{2 \pi R_S},
\end{equation}
after having set the speed of light to one i.e., $c=1$. Alternatively, the Lyapunov exponent of photon sphere can be written as
\begin{equation}
\lambda=\sqrt{\frac{f(r_{\rm ps})(2f(r_{\rm ps})-r_{\rm ps}^2 f''(r_{\rm ps})}{2 r_{\rm ps}^2}}.
\end{equation}

Although the relation (21) is not very accurate for small $l$, it is still helpful to investigate the relation between the QNMs and shadow radius ande their dependence on different physical quantities in a given black hole solution. As a particular example, we shall consider the effect of PFDM on QNMs then we make a simple with the effect of PFDM on shadow radius. The advantage of Eq. (21) relies on the fact then having determined the QNMs in the eikonal limit one can estimate the shadow radius or vice verse.

\section{QNMs of scalar field}
As a particular example let us consider the spherically symmetric black hole metric in PFDM given by \cite{Li,K1}
\begin{equation}
ds^2=-f(r)dt^2+\frac{dr^2}{f(r)}+r^2\left(d\theta^2+\sin^2\theta \,d\phi^2\right),
\end{equation}
with
\begin{equation}
f(r)=1-\frac{2M}{r}+\frac{k}{r}\ln\left(\frac{r}{|k|}\right),
\end{equation}
where $M$ is the black hole mass and $k$ is a parameter describing the intensity of the PFDM. If the PFDM is absent i.e., $k = 0$, the above space-time metric simply reduces to the Schwarzschild black hole. On the other hand, the massless scalar field in curved spacetime is described by the following equation
\begin{equation}
\frac{1}{\sqrt{-g}}\partial_{\mu}\left(\sqrt{-g} g^{\mu \nu} \partial_{\mu}\Phi  \right)=0.
\end{equation}

Involving a separation of variables the function $\Phi$ for the scalar field is given in terms of the spherical harmonics
\begin{equation}
\Phi(t,r,\theta,\phi)=\frac{1}{r}\,e^{-i\omega t}Y_{l}(r,\theta)\Psi(r),
\end{equation}
in which $l=0,1,2,...$  is known as the multipole number. After the separation of variables one can show that the field perturbation equation in the black hole spacetime is given by the Schrodinger wave-like equation
\begin{equation}
\frac{d^2\Psi}{dr_{\star}^2}+\left(\omega^2-V_S(r)\right)\Psi=0,
\end{equation}
where we have used the relation
\begin{equation}
dr_{\star}=\frac{dr}{f(r)}.
\end{equation}

Under the positive real part QNMs, by definition, satisfy the following boundary condition
\begin{equation}
\Psi(r_{\star})=C_{\pm}\,\exp\left( \pm i\,\omega \,r_{\star}  \right),\,\,\,\,\,r\to \pm \infty
\end{equation}
where $\omega$ can be written in terms of the real and imaginary part i.e., $\omega=\omega_{\Re}-i \omega_{\Im}$. In other words, we have  the real oscillation frequency and the imaginary part which is proportional to the decay rate of a given mode. The corresponding perturbation for the scalar field has the following effective potential 
\begin{eqnarray}\notag
V_S(r)&=&\left[1-\frac{2M}{r}+\frac{k}{r}\ln\left(\frac{r}{|k|}\right)\right]\\
&\times& \left[\frac{l(l+1)}{r^2} +\frac{2M+k-k\ln\left(\frac{r}{|k|}\right)}{r^3} \right].
\end{eqnarray}

 \begin{figure*}[ht]
\begin{center}
\includegraphics[width=8.7cm]{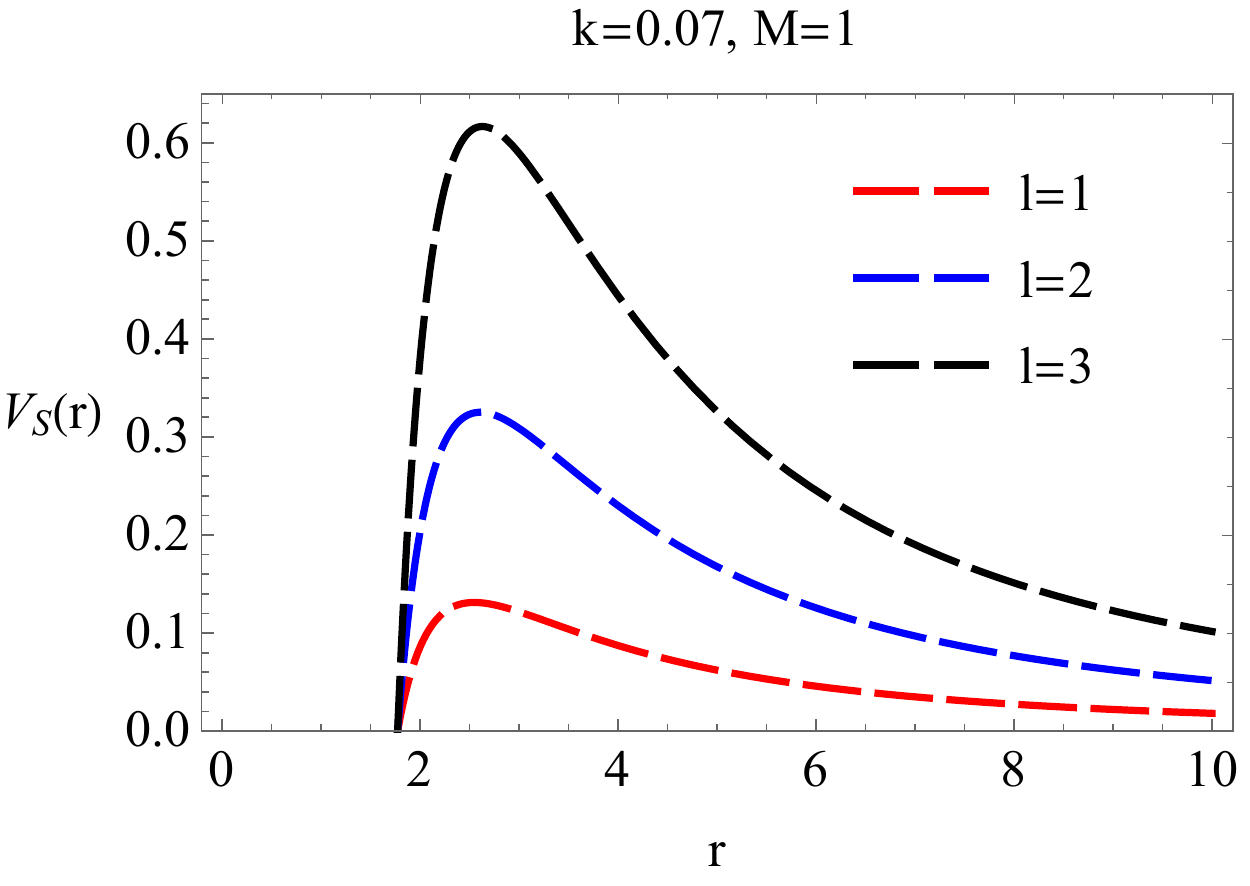}\;\;
\includegraphics[width=8.7cm]{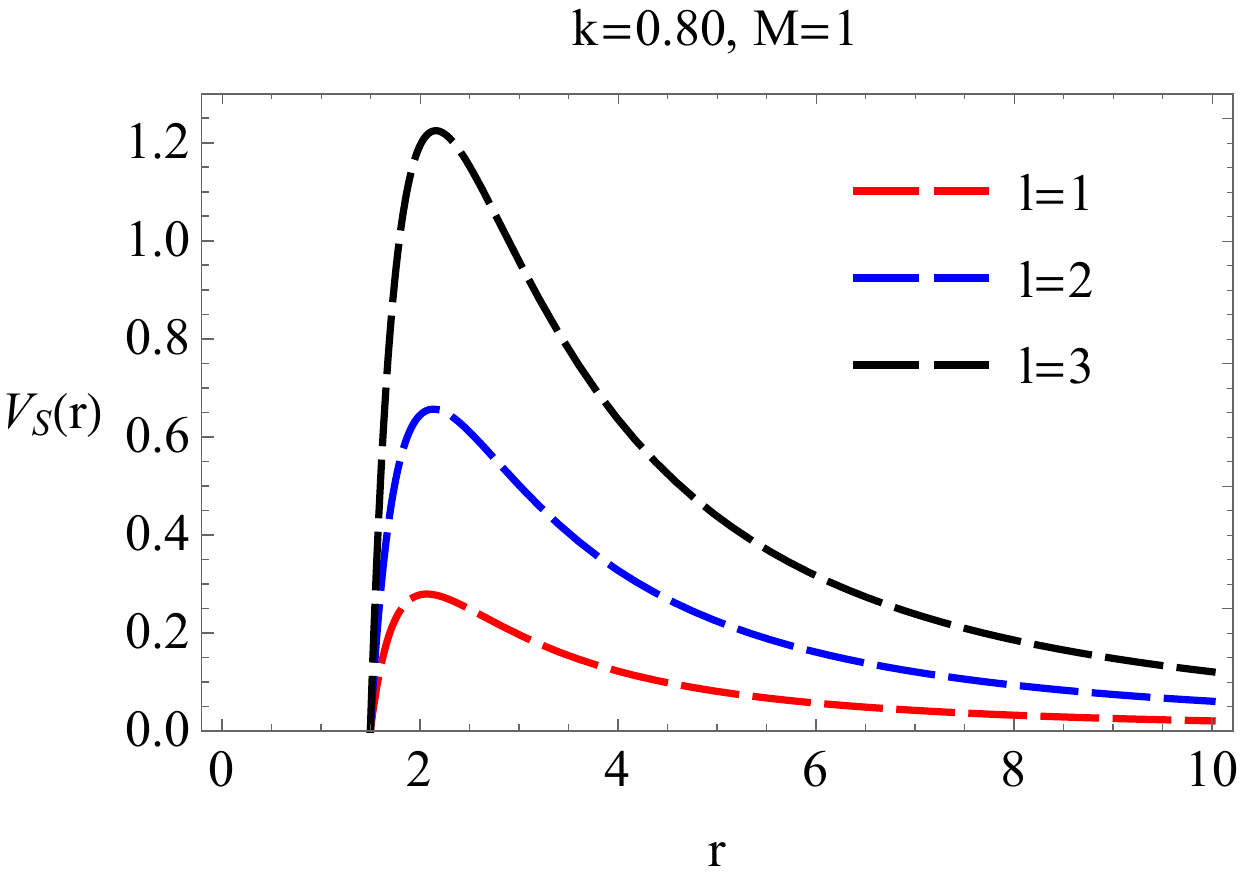}
\caption{The figures are the effective potentials of the scalar field perturbation $V_S$ for different values of PFDM parameter $k$. Changing the parameter $k$ changes the height of the potential barrier. }
 \end{center}
 \end{figure*}
 
 \begin{figure*}[ht]
\begin{center}
\includegraphics[width=8.7cm]{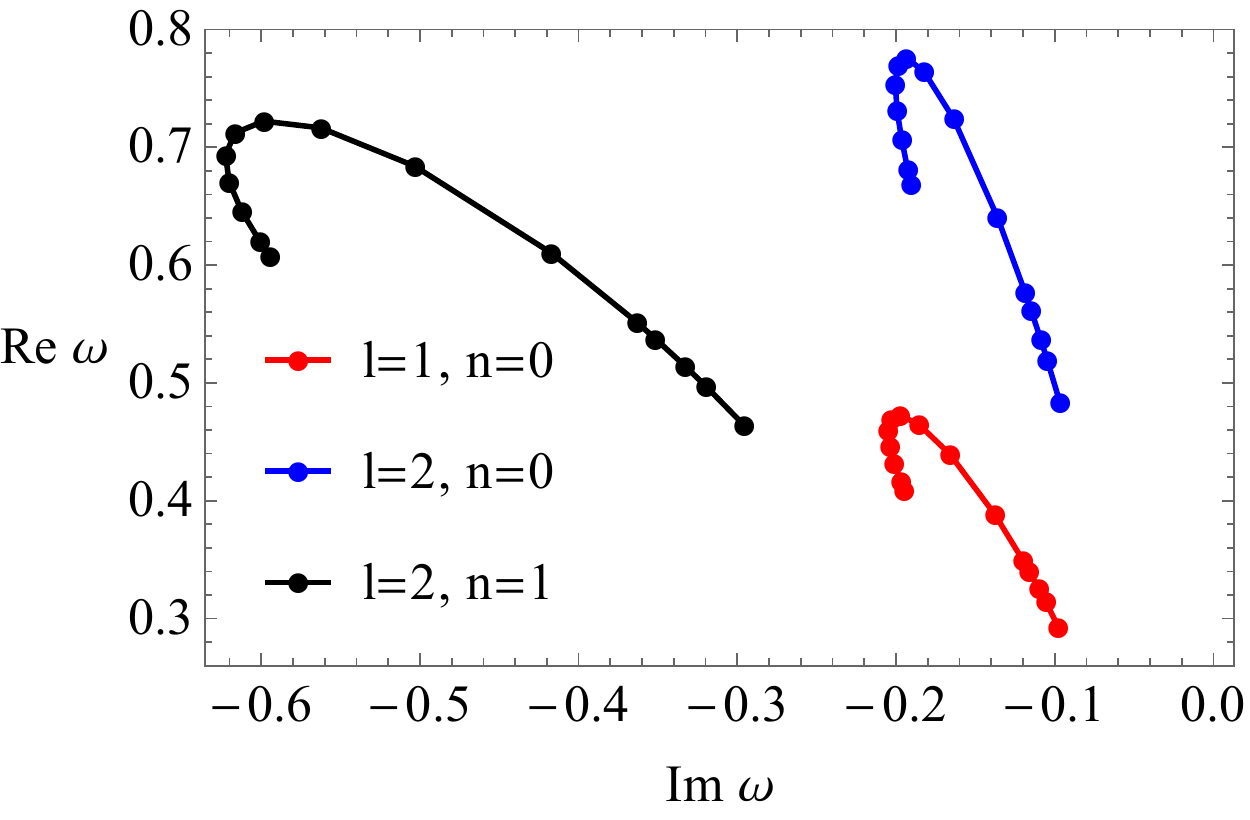}
\includegraphics[width=8.7cm]{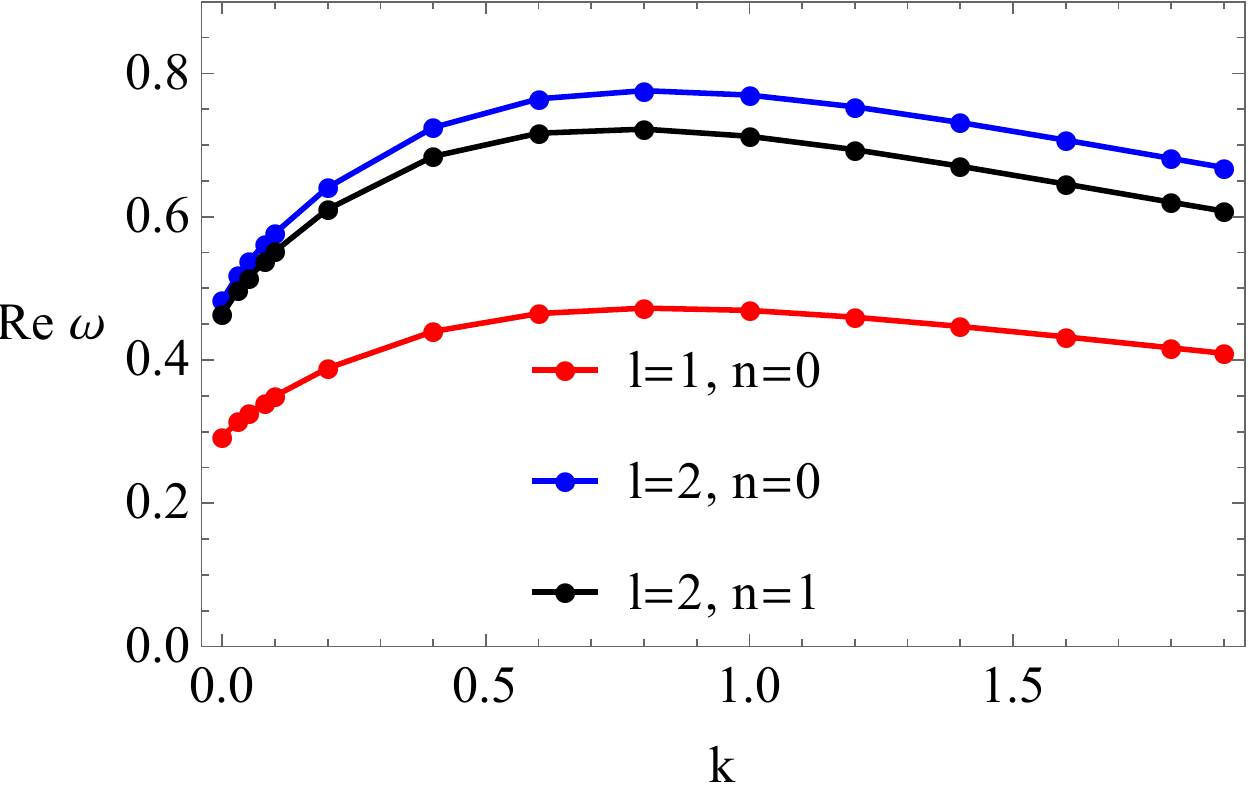}\;\;
\caption{Left panel: Dependence of the imaginary part of QNMs on $k$, for the case of scalar field. Right panel: Transmission coefficient for different $k$ in the case of scalar field.}
 \end{center}
 \end{figure*}

 \begin{figure*}[ht]
\begin{center}
\includegraphics[width=8.7cm]{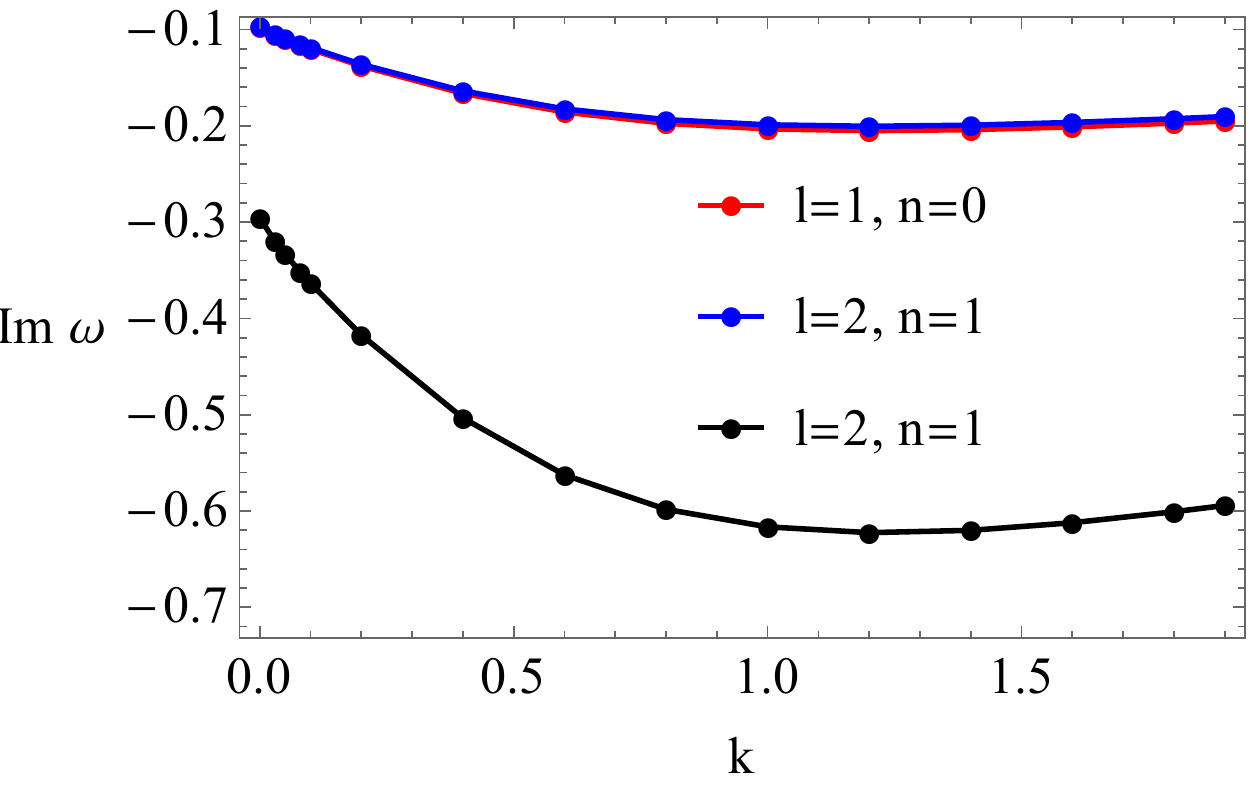}
\includegraphics[width=8.3cm]{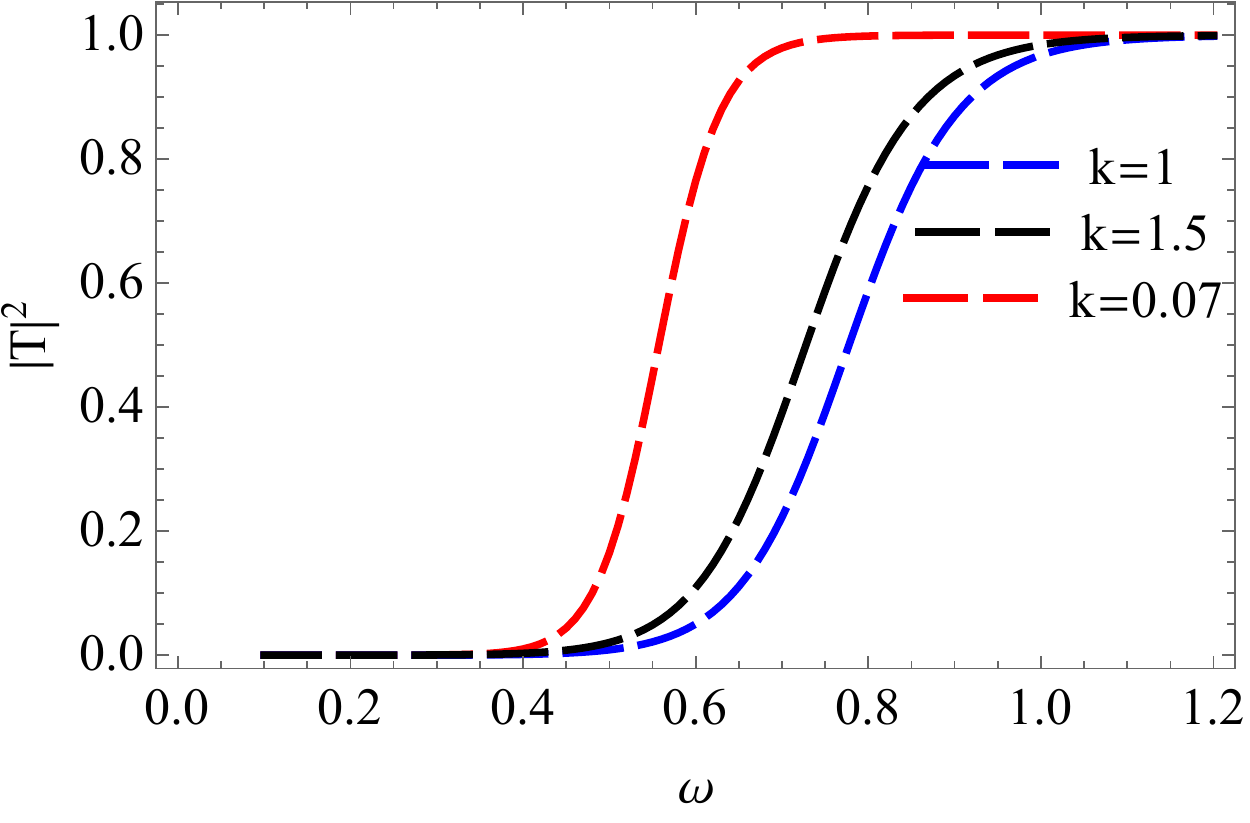}\;\;
\caption{The figures are the effective potentials of the scalar field perturbation $V_S$ for different values of PFDM parameter $k$. Dependence of the real and imaginary part of the QNM frequencies of the scalar field. Dependence of the real and imaginary part of the QNM frequencies of the scalar field. }
 \end{center}
 \end{figure*}

Having the relation for the effective potential we can use the WKB approach to compute the QNMs frequencies. The WKB method is based on the analogy with the problem of waves scattering near the peak
of the potential barrier in quantum mechanics, where $\omega$
plays a role of energy. The approach
was used by Schutz and Will \cite{Schutz}, developed to the third order by Iyer and Will \cite{{Iyer}}. 
%QNM frequencies in the WKB approximation, carried to third order beyond the eikonal approximation
%\begin{equation}
%\omega^2=\left[V_0+\sqrt{-2\,V_0''}\Lambda_2\right]-i\left(n+\frac{1}{2}\right)\sqrt{-2\,V_0''}(1+\Lambda_3)
%\end{equation}
%where
% \begin{eqnarray}\notag
%\Lambda_2 &=& \frac{1}{\sqrt{-2\,V_0''}}\Big\{\frac{1}{8}\Big(\frac{V_0^{(4)}}{V_0''}\Big)
%\Big(\frac{1}{4}+\alpha^2\Big)\\
%&-&\frac{1}{288}\Big(\frac{V_0^{(3)}}{V_0''}\Big)^2(7+60\alpha^2)\Big\},
%\nonumber\\\notag
%\Lambda_3 &=& \frac{1}{\sqrt{-2\,V_0''}}\Big\{\frac{5}{6912}\Big(\frac{V_0^{(3)}}{V_0''}\Big)^4
%\Big(77+188\alpha^2\Big)\\\notag
%&-&\frac{1}{384}\Big(\frac{V_0'''^2V_0^{(4)}}{V_0''^3}\Big)(51+100\alpha^2)
%\nonumber\\\notag
%&+&\frac{1}{2304}\Big(\frac{V_0^{(4)}}{V_0''}\Big)^2(67+68\alpha^2)\\
%&-&\frac{1}{288}\Big(\frac{V_0'''V_0^{(5)}}{V_0''^2}\Big)(19+28\alpha^2)\nonumber\\
%&-&\frac{1}{288}\Big(\frac{V_0^{(6)}}{V_0''}\Big)(5+4\alpha^2)\Big\},
%\end{eqnarray}
%where 
% \begin{eqnarray}
%\alpha=n+\frac{1}{2},\;\;V_0^{(m)}=\frac{d^mV}{dr_*^m}\Big|_{r_{\star}}.
%\end{eqnarray}
In the present paper we shall use the sixth order WKB approximation for calculating QNMs developed by Konoplya  \cite{KonoplyaWKB}.
%\begin{equation}
%i\frac{\omega_n^2-V_0}{\sqrt{-2\,V_0''}}-\sum_{i=2}^{6}\Lambda_i=n+\frac{1}{2}
%\end{equation}
%where the constants $\Lambda_4,\;\Lambda_5,\;\Lambda_6$.  Note that $V_0$ is the height and $V_0''$ is the second derivative with respect to the tortoise coordinate of the potential at the maximum. 
%The corrections depend on the value of the potential and higher derivatives of it at the maximum.
We estimate the values of the quasi-normal modes for the scalar perturbations given in Table 1.
\begin{table}[tbp]
\begin{tabular}{|l|l|l|l|l|}
\hline
 \multicolumn{1}{|c|}{ spin 0 } &  \multicolumn{1}{c|}{  $l=1, n=0$ } & \multicolumn{1}{c|}{  $l=2, n=0$ } & \multicolumn{1}{c|}{ $l=2, n=1$ }\\\hline
   $k$ & $\omega \,(WKB)$ &  $\omega \,(WKB)$ &  $\omega \,(WKB)$  \\ \hline
0 & 0.2929-0.0978 i & 0.4836-0.0968 i & 0.4638-0.2956 i  \\ 
0.03 & 0.3144-0.1058 i & 0.5190-0.1047 i & 0.4973-0.3199 i    \\
0.05 & 0.3255-0.1102 i & 0.5374-0.1090 i & 0.5146-0.3332 i  \\ 
0.08 & 0.3405-0.1163 i & 0.5620-0.1150 i & 0.5376-0.3517 i  \\ 
0.10 & 0.3497-0.1202 i & 0.5771-0.1188 i & 0.5517-0.3635 i \\ 
0.20 & 0.3886-0.1379 i& 0.6408-0.1362 i & 0.6103-0.4172 i\\ 
0.40 & 0.4398-0.1663 i& 0.7244-0.1638 i & 0.6841-0.5034 i\\ 
0.60 & 0.4648-0.1857 i& 0.7645-0.1825 i & 0.7165-0.5625 i  \\ 
0.80 & 0.4724-0.1974 i& 0.7759-0.1936 i & 0.7222-0.5983 i \\
1.00 & 0.4692-0.2032 i &0.7697-0.1990 i& 0.7122-0.6165 i \\ 
1.20 & 0.4597-0.2050 i & 0.7533-0.2005 i & 0.6935-0.6225 i   \\ 
1.40& 0.4468-0.2040 i & 0.7314-0.1994 i & 0.6704-0.6201 i   \\ 
1.60 & 0.4322-0.2011 i & 0.7068-0.1965 i & 0.6455-0.6122 i   \\ 
1.80 & 0.4169-0.1972 i  & 0.6813-0.1927 i & 0.6202-0.6008 i  \\ 
1.90 & 0.4093-0.1950 i & 0.6685-0.1904 i & 0.6077-0.5943 i  \\\hline
\end{tabular}
\caption{The real and imaginary parts of quasinormal frequencies of the scalar field in the background of BHPFDM with different dark matter parameters $k$. We find a reflecting point $k_0=0.80$ where the real part of QNMs reaches its maximal value, followed by an interval $k>k_0$, with a decrease of $\omega_{\Re}$. }
\end{table}

From Table I we see that by increasing $k$ the real part of QNMs increases reaching the maximal value at $k_0=0.80$. In fact, this is reflecting point, further increase of $k$ results with a decrease of the QNMs. But overall for any in the interval $k \in (0,2)$, the QNMs deviate from those of Schwarzschild BH resulting with greater values compared to the Schwarzschild black hole.  Moreover, from Fig. 2 we can see more clearly that the reflecting point $k_0$ is a generic feature for the real part of  QNMs frequencies having different $n$ and $l$, respectively. From Fig. 2 we also see that higher values of $\omega_{\Re}$ are obtained for the case $l=2$ and $n=1$. It's worth noting that we have not calculateted the QNMs for the the fundamental mode $l=n=0$ in Table I. This has to do with the fact that the WKB method is applicable when $l>n$ (where $n$ is the overtone number and $l$ is the multipole number) and does not give a satisfactory degree of precision for this fundamental mode.  Nevertheless, one can use the Frobenius method for more accurate calculation of this fundamental mode (see for example \cite{Konoplya:2018qov}).\\

\section{QNMs of electromagnetic field}
In this section we precede to study the effect of PFDM on the propagation of the electromagnetic field. To do so, we recall the wave equations for a test electromagnetic field,
\begin{equation}
\frac{1}{\sqrt{-g}}\partial_{\nu}\left[ \sqrt{-g} g^{\alpha \mu}g^{\sigma \nu} \left(A_{\sigma,\alpha} -A_{\alpha,\sigma}\right) \right]=0
\end{equation}
The four-potential $A_{\mu }$ can be expanded in
4-dimensional vector spherical harmonics as follows
\begin{eqnarray}\notag
A_{\mu }\left( t,r,\theta ,\phi \right) &=&\sum_{\ell ,m}\Big( \Big[
\begin{array}{c}
0 \\
0 \\
\frac{a(t,r)}{\sin \left( \theta \right) }\partial _{\phi }Y_{\ell
m}\left( \theta ,\phi \right) \\
-a\left( t,r\right) \sin \left( \theta \right) \partial _{\theta }Y_{\ell
m}\left( \theta ,\phi \right)%
\end{array}%
\Big] \\
&+& \Big[
\begin{array}{c}
f(t,r)Y_{\ell m}\left( \theta ,\phi \right) \\
h(t,r)Y_{\ell m}\left( \theta ,\phi \right) \\
k(t,r)\partial _{\theta }Y_{\ell m}\left( \theta ,\phi \right) \\
k(t,r)\partial _{\varphi }Y_{\ell m}\left( \theta ,\phi \right)%
\end{array}%
\Big] \Big) ,
\end{eqnarray}%
in which $Y_{\ell m}\left( \theta ,\phi \right) $ gives the spherical
harmonics. Note that the first term in the right-hand side has parity $\left(
-1\right) ^{\ell +1}$ (known as axial sector) and the second term
has parity $\left( -1\right) ^{\ell }$ (known as polar sector). If we simply
substitute this expansion into the Maxwell equations one can
find a second-order differential equation for the radial part as (see \cite%
{Cosimo}\ for details of calculations)%
\begin{equation}
\frac{d^{2}\Psi \left( r_{\ast }\right) }{dr_{\ast }^{2}}+\left[ \omega
^{2}-V_{E}\left( r_{\ast }\right) \right] \Psi \left( r_{\ast }\right) =0,
\end{equation}%
for both axial and polar sectors, and $r_{\ast }=\int f^{-1}\left( r\right)
dr$\ being the tortoise coordinate. The mode $\Psi \left( r_{\ast }\right) $
is a linear combination of the functions $a(t,r)$, $f(t,r)$, $h(t,r)$, and $%
k(t,r)$, but a different functional dependence based on the parity; for
axial sector the mode is given by 
\begin{equation}
a(t,r)=\Psi \left( r_{\ast }\right) 
\end{equation}

whereas for polar sector it is 
\begin{equation}
\Psi \left( r_{\ast }\right) =\frac{r^{2}}{%
\ell (\ell +1)}\left[ \partial _{t}h(t,r)-\partial _{r}f(t,r)\right].
\end{equation}
The corresponding  effective potential in our case is found to be
\begin{equation}
V_E(r)=\left[1-\frac{2M}{r}+\frac{k}{r}\ln\left(\frac{r}{|k|}\right)\right]\,\frac{l(l+1)}{r^2}.
\end{equation}

\begin{figure*}[ht]
\begin{center}
\includegraphics[width=8.4cm]{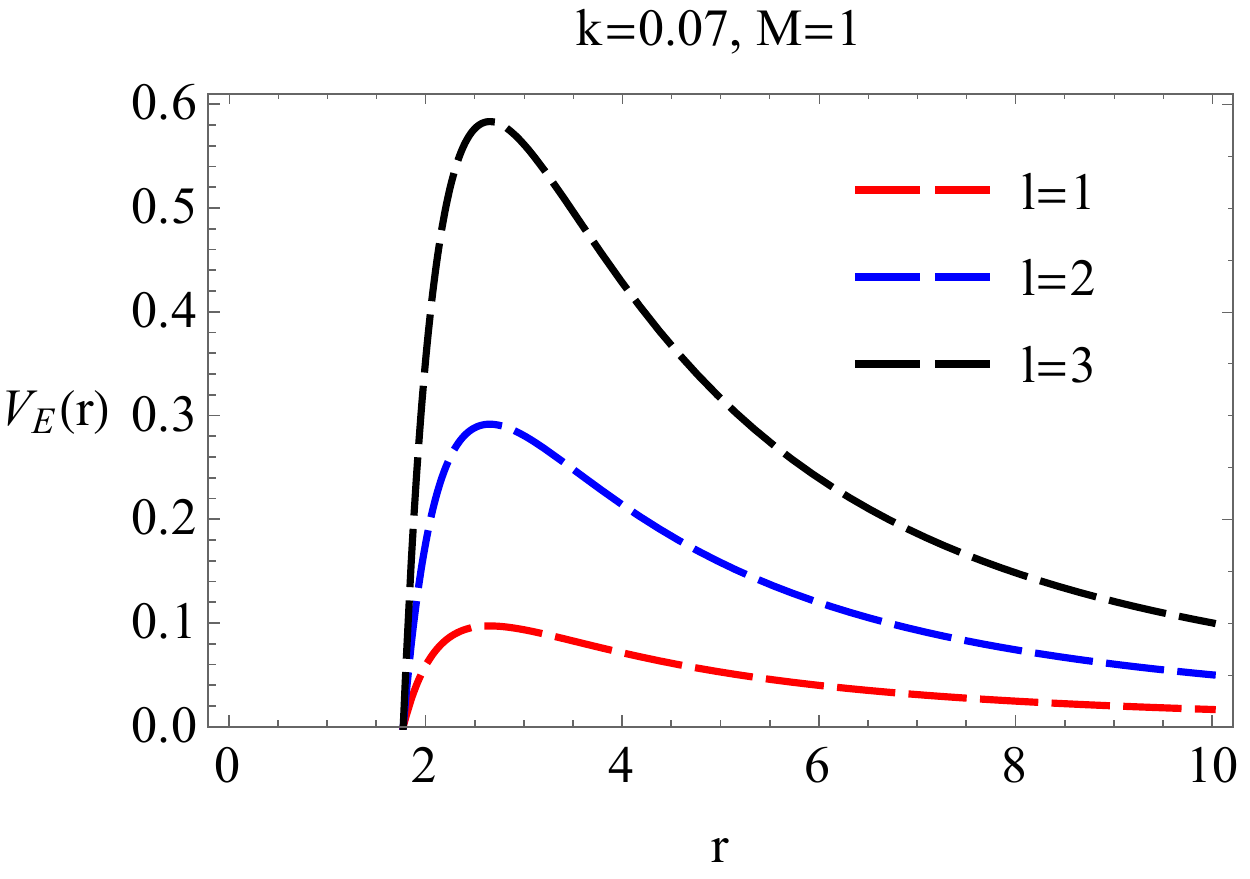}\;\;
\includegraphics[width=8.4cm]{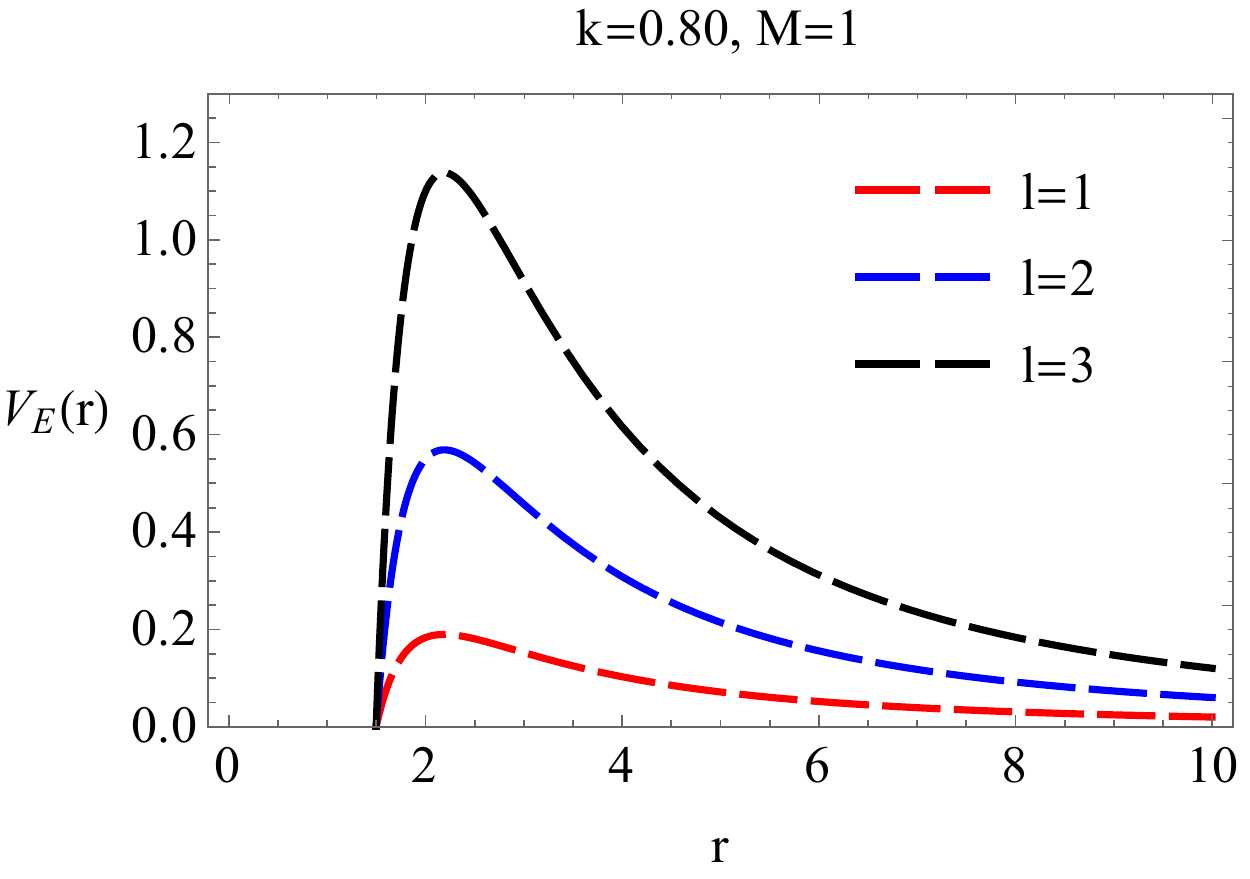}
\caption{The effective potentials of electromagnetic field perturbation $V_E$ near black hole with PFDM with $(M=1)$ and different $k$.  }
 \end{center}
 \end{figure*}
 
 \begin{figure*}[ht]
\begin{center}
\includegraphics[width=8.7cm]{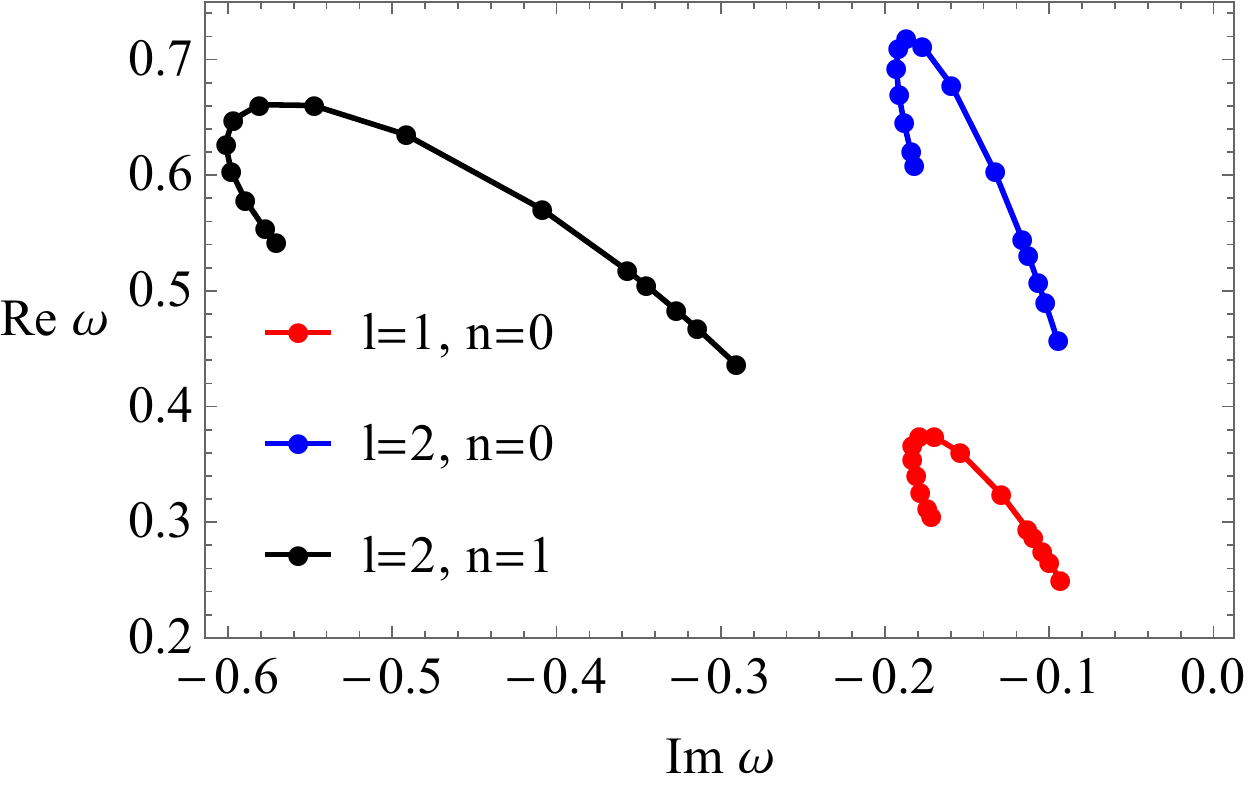}\;\;
\includegraphics[width=8.7cm]{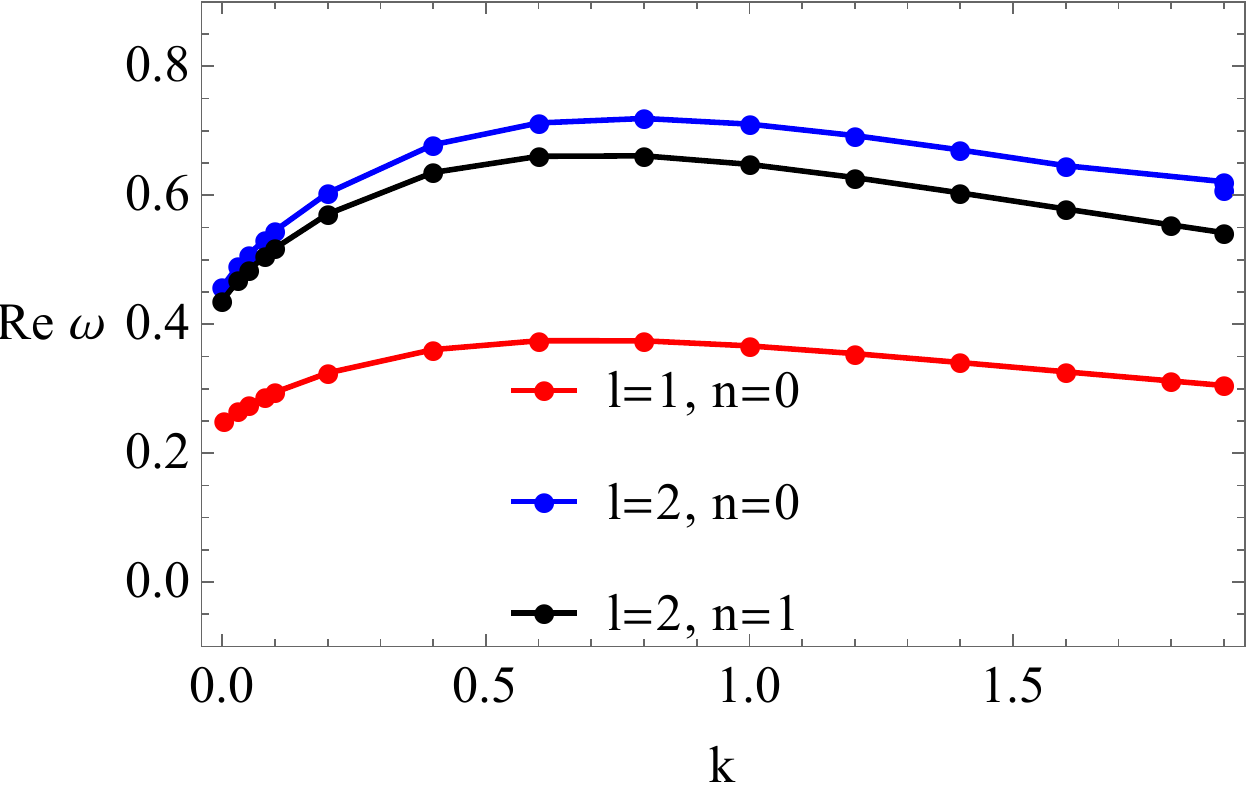}
\caption{Left panel: Dependence of the real part of QNM frequencies on the imaginary part for electromagnetic test field. Right panel: Dependence of the real part of QNM frequencies on the PFDM parameter $k$. As in the case of scalar field, a reflecting point is observed at $k_0=0.80$. }
 \end{center}
 \end{figure*}
 
 \begin{figure*}[ht]
\begin{center}
\includegraphics[width=8.4cm]{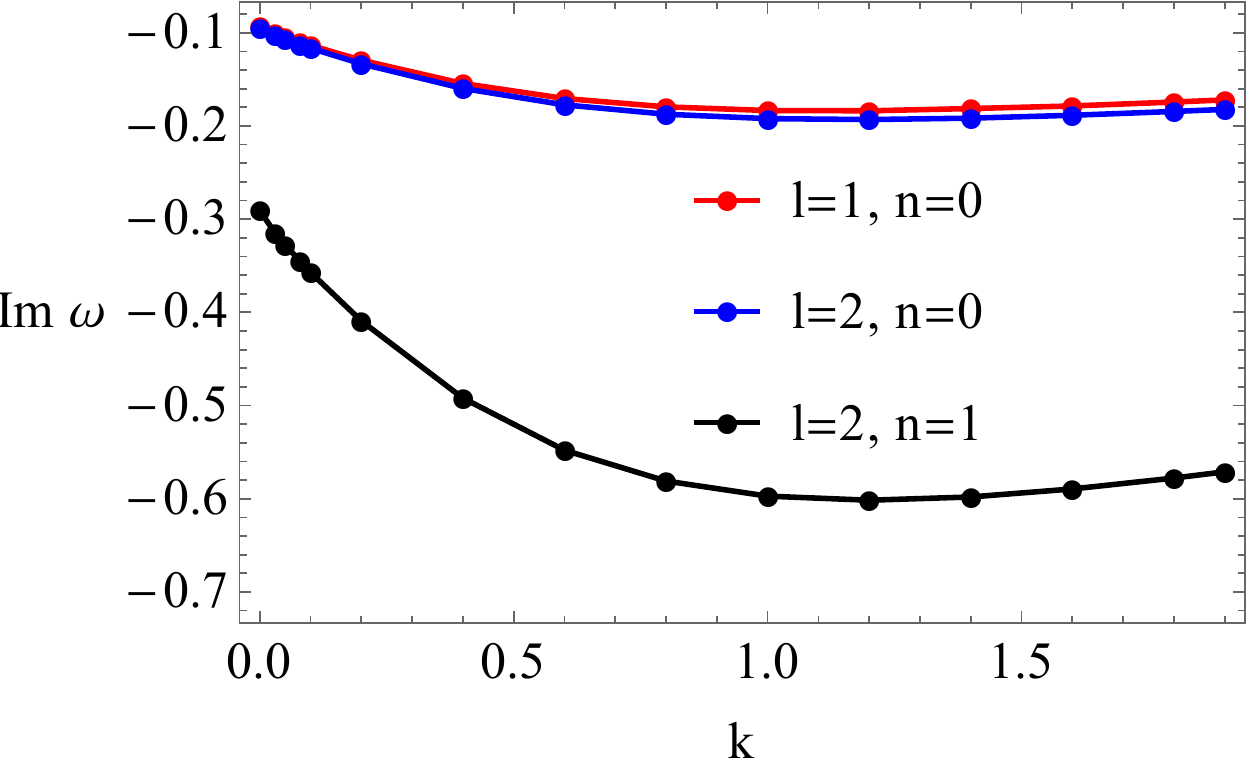}\;\;
\includegraphics[width=8.1cm]{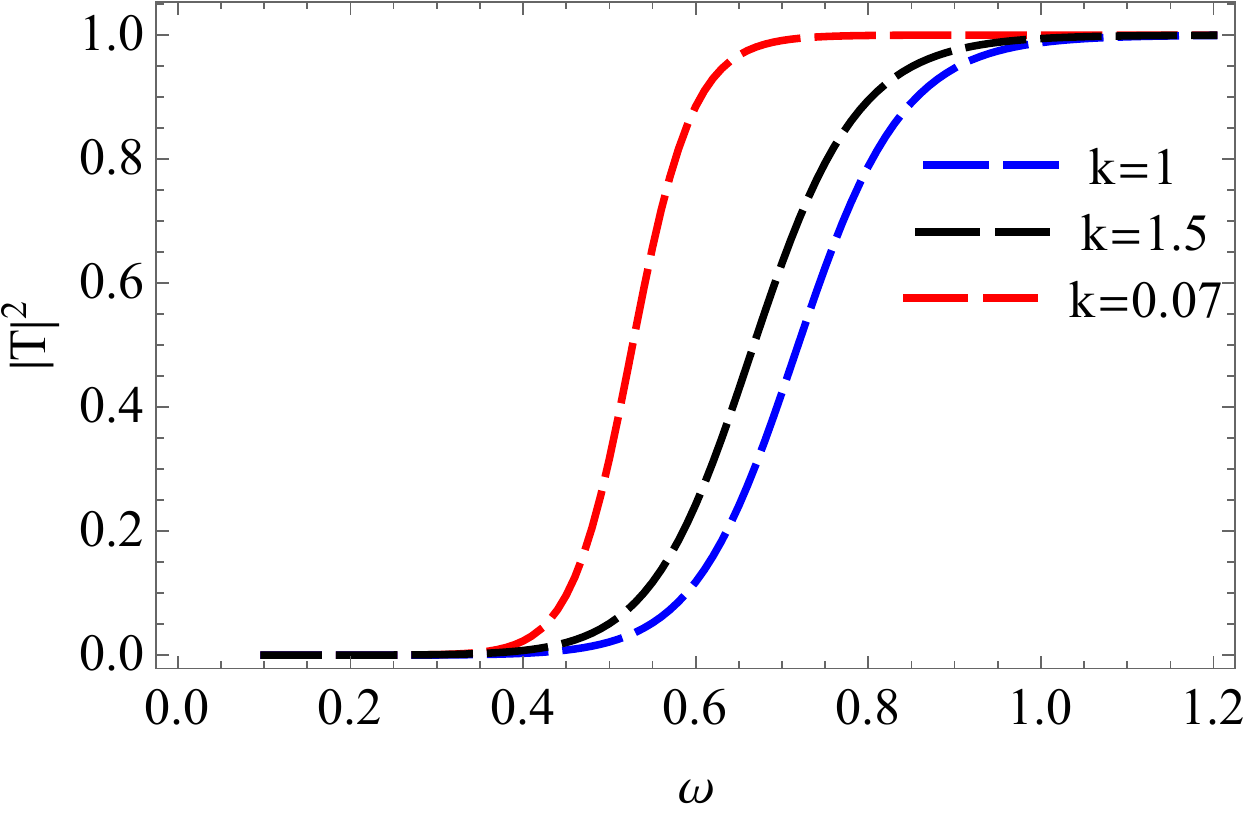}
\caption{Left panel: Dependence of the imaginary part of QNM frequencies the PFDM parameter $k$. Right panel: Dependence of the transmission coefficient for electromagnetic test field.  }
 \end{center}
 \end{figure*}

\begin{table}[tbp]
\begin{tabular}{|l|l|l|l|l|}
\hline
 \multicolumn{1}{|c|}{ spin 1 } &  \multicolumn{1}{c|}{  $l=1, n=0$ } & \multicolumn{1}{c|}{  $l=2, n=0$ } & \multicolumn{1}{c|}{ $l=2, n=1$ }\\\hline
   $k$ & $\omega \,(WKB)$ & $\omega \,(WKB)$ & $\omega \,(WKB)$   \\ \hline
0 & 0.2482-0.0926 i & 0.4576-0.0950 i & 0.4365-0.2907 i  \\ 
0.03 & 0.2659-0.1001 i & 0.4907-0.1027 i & 0.4677-0.3145 i    \\
0.05 & 0.2750-0.1042 i & 0.5079-0.1070 i & 0.4836-0.3274 i  \\ 
0.08 & 0.2870-0.1098 i & 0.5308-0.1128 i & 0.5049-0.3455 i  \\ 
0.10 & 0.2943-0.1134 i & 0.5448-0.1165 i & 0.5177-0.3570 i \\ 
0.20 & 0.3243-0.1294 i& 0.6034-0.1333 i & 0.5708-0.4091 i\\ 
0.40 & 0.3605-0.1544 i& 0.6782-0.1597 i & 0.6352-0.4920 i\\ 
0.60 & 0.3743-0.1705 i& 0.7120-0.1773 i & 0.6603-0.5479 i  \\ 
0.80 & 0.3742-0.1795 i& 0.7190-0.1874 i & 0.6610-0.5810 i \\
1.00 & 0.3663-0.1834 i &0.7102-0.1921 i& 0.6478-0.5972 i \\ 
1.20 & 0.3543-0.1837 i & 0.6924-0.1931 i & 0.6273-0.6016 i   \\ 
1.40& 0.3405-0.1813 i & 0.6701-0.1916 i & 0.6035-0.5981 i   \\ 
1.60 & 0.3261-0.1785 i & 0.6457-0.1885 i & 0.5786-0.5895 i   \\ 
1.80 & 0.3119-0.1743 i  & 0.6209-0.1844 i & 0.5539-0.5778 i  \\ 
1.90 & 0.3049-0.1720 i & 0.6086-0.1822 i & 0.5418-0.5711 i  \\\hline
\end{tabular}
\caption{The real and imaginary parts of quasinormal frequencies of the electromagnetic field in the background perfect fluid black hole with different $k$. A reflecting point is found at $k_0$.}
\end{table}

Firstly, from Fig. 4 we see that the effective potential is strongly affected by the PFDM parameter $k$. For example, we observe higher values of the potential barrier when $k \simeq 0.80$. Using the WKB approximation to the sixth order we compute the real and imaginary part of QNMs frequencies \cite{KonoplyaWKB}. 
In Table II, we provide the values of QNMs having $l=1,2,3$ and $n=0,1$, respectively. We see a similar effect of $k$, namely when $k$ increases the real part of QNMs increases and reaches the maximum at $k \simeq 0.8$. This reflecting point can be seen also from Fig. 5 where maximal values for the real part are obtained at $k_0$. The values of QNMs having $l=2$ and $n=1$ are slightly higher. In addition, from Table I and Table II we see that the values of QNMs for the electromagnetic field are slightly smaller compared to the scalar field.

\section{Scattering and greybody factor}
Greybody factors are important quantities to determine for example the amount of the initial quantum radiation in the vicinity of the black hole horizon which
is reflected back to it by the potential barrier. Moreover we can use the Hawking equation to estimate the amount of radiation which will reach the observer located at infinity. Starting from the Schrodinger-like equation which describes the scattering of waves in the regular BH spacetimes and in has the asymptotic solutions 
\begin{equation}
\Psi = A\exp(-i \omega r_{\star})+B\exp(i \omega r_{\star}),\,\,\,\,r_{\star} \to -\infty,
\end{equation}
\begin{equation}
\Psi =C\exp(-i \omega r_{\star})+D\exp(i \omega r_{\star}),\,\,\,\,r_{\star} \to +\infty ,
\end{equation}
where we have to impose the following relations: for waves incoming towards the BHPFDM from infinity
we have $B= 0$, while for the reflection amplitude
$R = D/C$. Finally for the transmission
amplitude we can write $T  = A/C$. In particular the last equation can be written as
\begin{equation}
\Psi = T\exp(-i \omega r_{\star})\,\,\,\,r_{\star} \to -\infty ,
\end{equation}
\begin{equation}
\Psi =\exp(-i \omega r_{\star})+R\exp(i \omega r_{\star}),\,\,\,\,r_{\star} \to +\infty.
\end{equation}

Next we need to compute the square of the amplitude of the wave function which is partially transmitted and partially reflected by the potential barrier. The total probability of finding
this wave in the whole region should give
\begin{equation}
|R|^2+|T|^2=1.
\end{equation}

The most interesting case to study the greybody factor is the case having $\omega \simeq V(r_0)$.
Namely, we can the WKB approximation to calculate the transmission and
reflection coefficients. In particular the reflection amplitude is given by
\begin{equation}
R=\frac{1}{\sqrt{1+\exp(-2 \pi i K)}},
\end{equation}
where 
\begin{equation}
K=i\frac{\omega_n^2-V(r_0)}{\sqrt{-2\,V''(r_0)}}-\sum_{i=2}^{6}\Lambda_i.
\end{equation}

Using (42) we can find an expression for the
transmission coefficient given by
\begin{equation}
|T|^2=1-|\frac{1}{\sqrt{1+\exp(-2 \pi i K)}}|^2.
\end{equation}

In Figures 3 and 6 we show the dependence of 
the transmission coefficients on $k$ for the scalar,
electromagnetic and  fields, respectively.

\section{Shadow radius of black hole in PFDM}
\begin{figure*}[ht]
\begin{center}
\includegraphics[width=8.3cm]{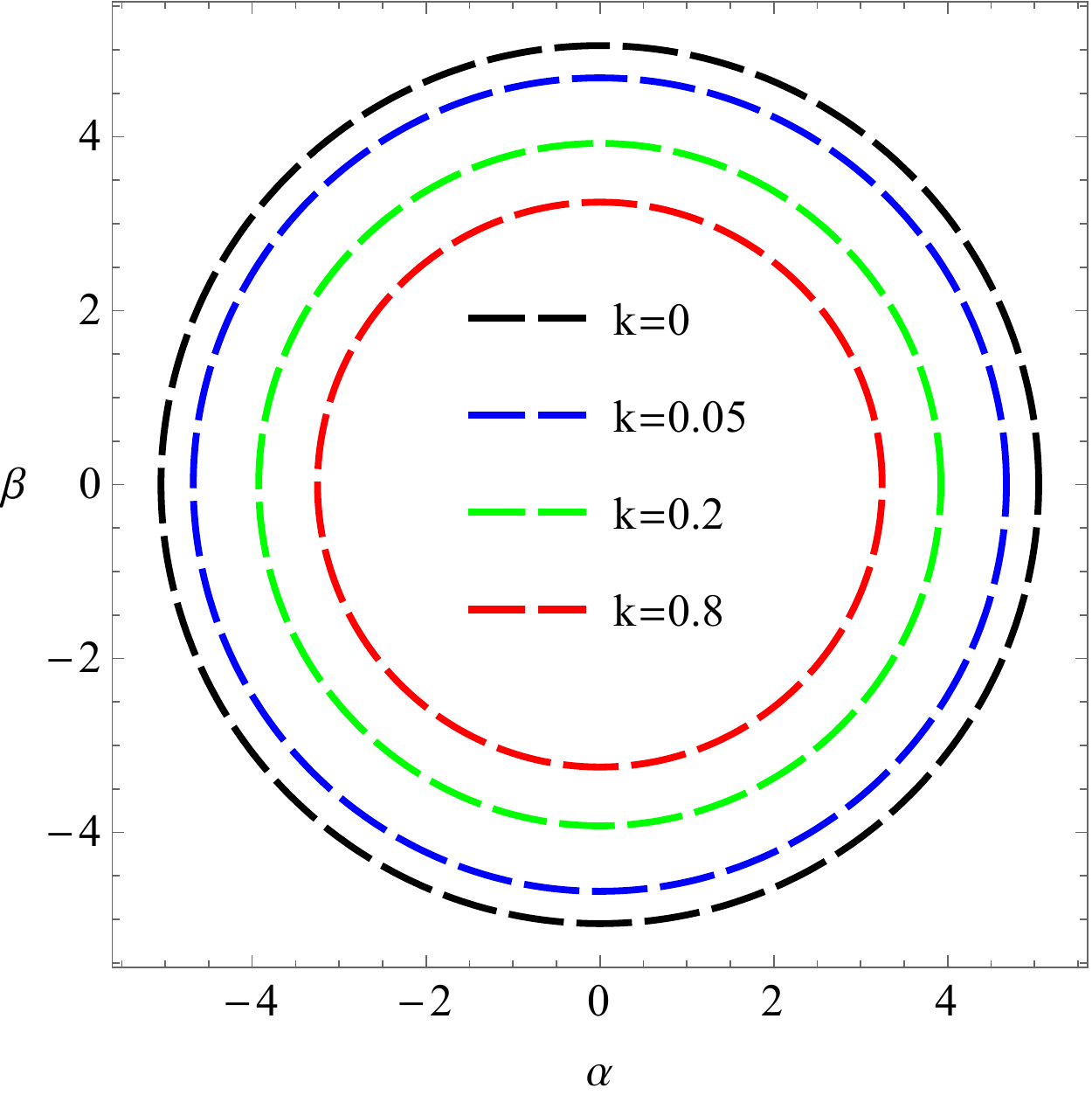}
\includegraphics[width=7.8 cm]{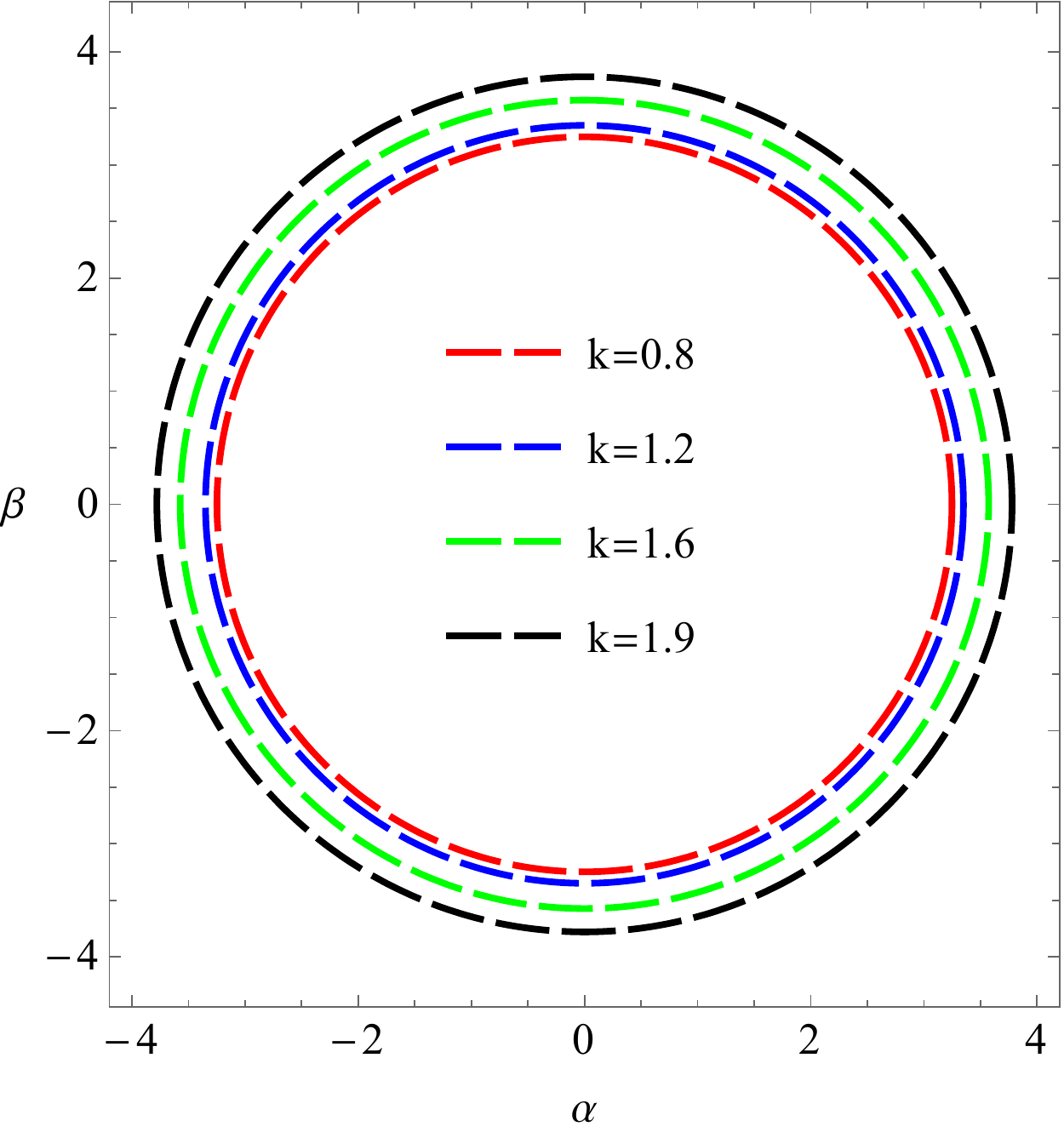}
\caption{Shape of shadows for different $k$ and $M=1$. We observe that there exists a reflecting point $k_0=0.8$. Due to the fact that we the black hole mass is fixed by changing the PFDM parameter we decrease the event horizon which leads to a smaller radius of the shadow. }
 \end{center}
 \end{figure*}
 
In general, it is well known that the shadow shape depends on whether the rotation of the black hole is considered.
For static and spherical symmetric black holes, the shadow has spherical symmetry as well and is described by the photon sphere. In this case, the above conditions (\ref{condition}) can be simplified into,
\begin{equation}\lb{PSradius}
2 - \frac{rf'(r)}{f(r)}=0.
\end{equation}
with
\begin{equation}
f(r)=1-\frac{2M}{r}+\frac{k}{r}\ln\left(\frac{r}{|k|}\right).
\end{equation}
The solution of this equation determines the radius $r_{\rm ps}$ of the photon sphere is found to be
\begin{equation}\lb{rps2}
r_{\rm ps} = \frac{3}{2}\text{LambertW}\left[ \frac{2}{3}\exp\left( \frac{6M+k}{3k} \right) \right].
\end{equation}
and shadow radius
\begin{equation}\lb{RS}
R_{\rm s} =\frac{ \frac{3}{2}\text{LambertW}\left[ \frac{2}{3}\exp\left( \frac{6M+k}{3k} \right) \right]}{ \sqrt{1-\frac{2M}{ \frac{3}{2}\text{LambertW}\left[ \frac{2}{3}\exp\left( \frac{6M+k}{3k} \right) \right]}+\mathcal{F}}}.
\end{equation}
where
\begin{eqnarray}\notag
\mathcal{F}&=&\frac{k}{ \frac{3}{2}\text{LambertW}\left[ \frac{2}{3}\exp\left( \frac{6M+k}{3k} \right) \right]}\\
&\times& \ln\left[ \frac{ \frac{3}{2}\text{LambertW}\left[ \frac{2}{3}\exp\left( \frac{6M+k}{3k} \right) \right]}{|k|} \right]
\end{eqnarray}
\begin{figure}[ht]
\begin{center}
\includegraphics[width=8.6cm]{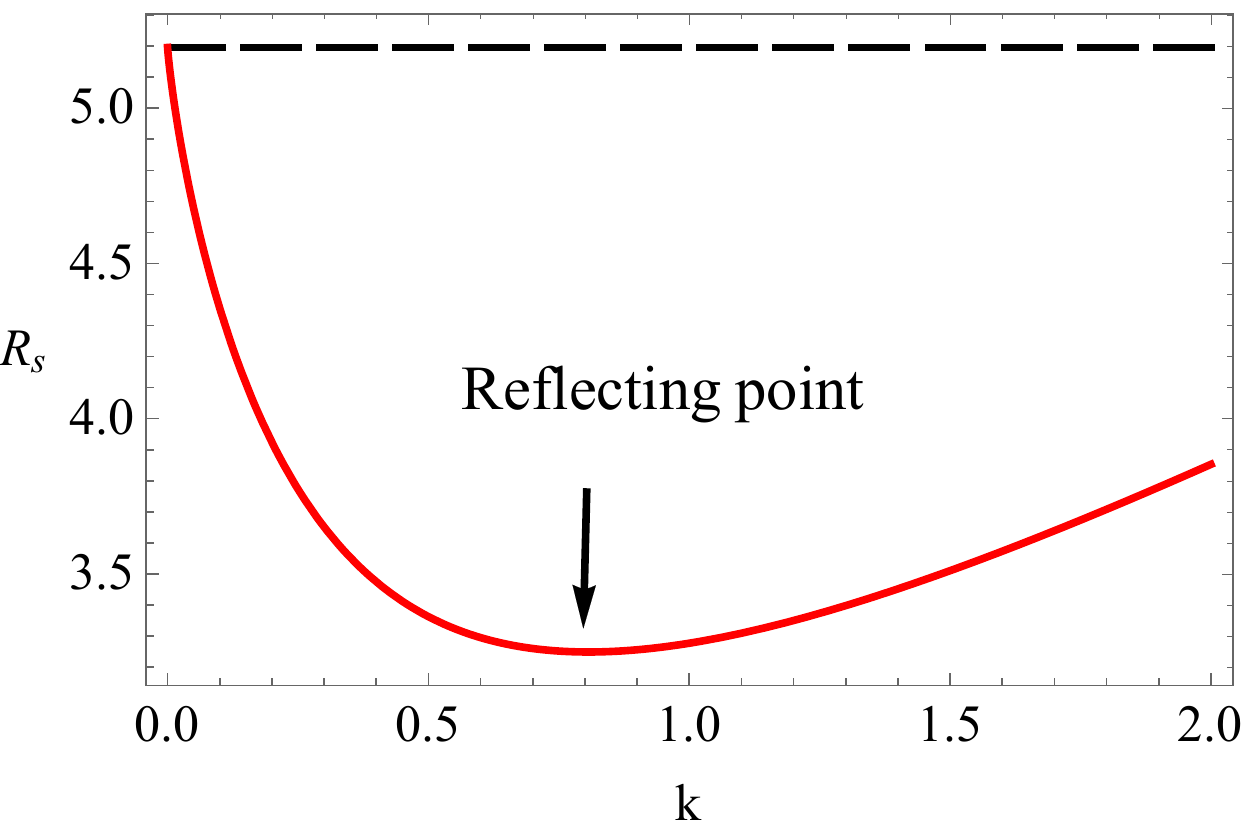}
\caption{Shadow radius against the PFDM parameter $k$ (red curve). We see that there exist a reflecting point $k_0 \simeq 0.8$.  The black curve corresponds to the shadow radius for the Schwarzschild vacuum BH $R_s=3 \sqrt{3}$.}
 \end{center}
 \end{figure}
\begin{table}[tbp]
\begin{tabular}{|l|l|}
\hline
   $k$ in unites of $M$   & $R_S$ in unites of $M$   \\ \hline
0 & 3 $\sqrt{3}\simeq$ 5.196  \\ 
0.03 & 4.8425   \\ 
0.05 &  4.6773  \\ 
0.08 & 4.4724   \\ 
0.10 & 4.3558 \\
0.20 & 3.9241  \\  
0.40 &  3.4744  \\ 
0.60 & 3.2943   \\ 
0.80 &  3.2480   \\ 
1.00 &  3.2762 \\
1.20 &  3.3493 \\  
1.40 &   3.4512 \\ 
1.60 &   3.5726 \\ 
1.80 &   3.7076 \\ 
1.90 & 3.7791 \\ 
 \hline
\end{tabular}
\caption{The size of shadow radius for different values of $k$. We observe a reflecting point near $k_0=0.8$. }
\end{table}

In Table III, we observe a decrease of the shadow radius in the interval $k<k_0$ while reaching its minimal value at $k=k_0$. This reflecting point can be seen also from Fig. 8 and the corresponding shadow images given by Fig. 7. The decrease of shadow radius is a consequence of the decrease of horizon radius, namely we keep the black hole mass fix and change $k$. Using the expression for the shadow radius (\ref{RS}) we can determine the reflection point by simply taking its derivative and solve for $k$. We find
\begin{equation}
k_0=\frac{3}{1+e}\simeq 0.80.
\end{equation}

Thus we found the exactly same reflection point as in the case of QNMs. Considering the inverse relation between $R_S$ and the real part of gravitational QNMs
\begin{equation}
\omega_{\Re} = \lim_{l \gg 1} \frac{l}{R_S},
\end{equation}
we can explain why the reflecting point $k_0$ yields contrary results for the shadow radius compared to QNMs. In other words, contrary to QNMs, the shadow radius decreases in the interval $k<k_0$ reaching its minimal vale at $k_0$. Further increase of $k$ in the interval $k>k_0$, suggest an increase of the shadow radius. We note that this result is consistent with the one obtained by Hou at el. \cite{Hou:2018avu} where authors studied shadow of rotating black holes in PFDM.  From a physical point of view, this can be explained by the fact that a decrease of the black hole horizon leads to a decrease of the shadow radius and consequently higher values for the oscillation frequencies.

\section{Conclusion}
In this paper we have pointed out a simple connection between the real part of the QNMs and shadow radius in the eikonal limit related by Eq. (21) which is accurate only for large values of $l$. Note that even in the limit $l\gg 1$ this correspondence is not guaranteed for gravitational fields, where the link between the null geodesics and quasinormal modes is shown to be violated in the Einstein-Lovelock theory \cite{Konoplya:2017wot}.  We have performed a detailed analyses of QNMs for a massless scalar and electromagnetic field perturbations in a spacetime background of black holes surrounded by PFDM.  Using the sixth-order WKB approximation we find that the QNM frequencies of black holes in presence of PFDM strongly depends upon the perfect fluid dark matter parameter $k$. These results show that QNM frequencies in the future can play a significant role as an indirect way of detecting dark matter near the black hole. We have found that for $k>0$, the real part and the absolute value of the imaginary part of QNM frequencies increases suggesting that the field perturbations in BHPFDM decays more rapidly compared to Schwarzschild vacuum BH.

More importantly, it is shown that there exists a reflecting point $k_0$ corresponding to maximal values for the real part of QNM frequencies.  That is, if the PFDM parameter $k$ increases in the interval $k<k_0$, the real part of QNMs also increases reaching maximum value at $k=k_0$. In the final part of this work we argued that $k_0$ is a reflecting point for the shadow radius of BHPFDM and corresponds to its minimal value.  Using the inverse relation between the real part of QNMs and shadow radius we can easily obtain the above conclusion without computing the relevant shadow radius. In the case of small values of $l$, the results depend on the spin of the test field, however Eq. (21) can still be useful to investigate the dependece of shadow radius on physical quantities such as the PFDM parameter $k$ in our case. Finally, we note that a correspondence between the Hawking radiation and QNMs was explored in Refs. \cite{Hod}-\cite{corda2}, and the relation between black hole shadow and the black hole thermodynamics was used recently in Ref. \cite{Zhang:2019glo} to study the thermodynamics phase structure of the black hole. This suggests that QNMs can encode valuable information about the thermodynamics phase structure or the stability of the black hole.

\end{document}